\documentclass[10pt,aps,prl,twocolumn,superscriptaddress,float,amsmath]{revtex4-2}
\usepackage{xcolor}
\usepackage{amsmath,bm,epsfig}
\definecolor{g-blue}{rgb}{0.83,0.95,1}
\usepackage{bm}
\usepackage{amssymb}
\usepackage{siunitx}
\usepackage{amsfonts}
\usepackage{amsmath}
\usepackage{mathtools}
\usepackage{ulem}
\usepackage{upgreek} % µ in Roman in math environment: \upmu
\usepackage{textcomp} % µ in Roman: \textmu
\usepackage{multirow}
\usepackage{booktabs} % Add this to the preamble if not already present
\usepackage[utf8]{inputenc}
\usepackage{graphicx}
\usepackage{hyperref}
\hypersetup{colorlinks=true,allcolors=blue,breaklinks=true}
\usepackage[all]{hypcap}

\def\vec#1{{\bm{#1}}}

\begin{document}
\title{Nanoscale spin-wave frequency-selective limiter for 5G technology}

\author{Kristýna Davídková}
    \email[Email address: ]{kristyna.davidkova@univie.ac.at}% Your name
    \affiliation{University of Vienna, Faculty of Physics, Boltzmanngasse 5, Vienna, Austria.}
    \affiliation{Vienna Doctoral School in Physics, University of Vienna, Boltzmanngasse 5, Vienna, Austria.}
    
\author{Khrystyna Levchenko}
    \affiliation{University of Vienna, Faculty of Physics, Boltzmanngasse 5, Vienna, Austria.}

\author{Florian Bruckner}
  \affiliation{University of Vienna, Faculty of Physics, Boltzmanngasse 5, Vienna, Austria.}

\author{Roman~Verba}
    \affiliation{Institute of Magnetism, 36-b Vernadskogo blvd., Kyiv 03142, Ukraine.}

\author{Fabian Majcen}
    \affiliation{University of Vienna, Faculty of Physics, Boltzmanngasse 5, Vienna, Austria.}
    \affiliation{Vienna Doctoral School in Physics, University of Vienna, Boltzmanngasse 5, Vienna, Austria.}

\author{Qi Wang}
    \affiliation{School of Physics, Hubei Key Laboratory of Gravitation and Quantum Physics, Institute for Quantum Science and Engineering, Huazhong University of Science and Technology, Wuhan, China.}

\author{Morris~Lindner}
    \affiliation{INNOVENT e. V. Technologieentwicklung, Prüssingstraße 27 B, Jena, Germany.}
    
\author{Carsten~Dubs}
    \affiliation{INNOVENT e. V. Technologieentwicklung, Prüssingstraße 27 B, Jena, Germany.}

\author{Vincent~Vlaminck}
    \affiliation{IMT Atlantique, Lab-STICC—UMR 6285 CNRS, Technopole Brest-Iroise CS83818, Brest, France.}

\author{Jan Klíma}
    \affiliation{Brno University of Technology, Faculty of Mechanical Engineering, Technická 2, Brno, Czech Republic.}

\author{Michal Urbánek}
    \affiliation{Brno University of Technology, Faculty of Mechanical Engineering, Technická 2, Brno, Czech Republic.}
    \affiliation{CEITEC BUT, Brno University of Technology, Purkyňova 123, Brno, Czech Republic.}

\author{Dieter Suess}
  \affiliation{University of Vienna, Faculty of Physics, Boltzmanngasse 5, Vienna, Austria.}

  \author{Andrii Chumak}
    \email[Email address: ]{andrii.chumak@univie.ac.at}% Your name
  \affiliation{University of Vienna, Faculty of Physics, Boltzmanngasse 5, Vienna, Austria.}
    
\date{\today} % Leave empty to omit a date

\begin{abstract}
 Power limiters are essential devices in modern radio frequency (RF) communications systems to protect highly sensitive input channels from large incoming signals. Nowadays-used semiconductor limiters suffer from high electronic noise and switching delays when approaching the GHz~range, which is crucial for the modern generation of 5G communication technologies aiming to operate at the EU 5G high band (24.25--27.5~GHz). The proposed solution is to use ferrite-based Frequency Selective Limiters (FSLs), which maintain their efficiency at high GHz frequencies, although they have only been studied at the macroscale so far.
In this study, we demonstrate a proof of concept of nanoscale FSLs. The devices are based on spin-wave transmission affected by four-magnon scattering phenomena in a 97-nm-thin Yttrium Iron Garnet (YIG) film. Spin waves were excited and detected using coplanar waveguide (CPW) transducers of the smallest feature size of $250\,\mathrm{nm}$. The FSLs are tested in the frequency range up to $25\,\mathrm{GHz}$, and the key parameters are extracted (power threshold, power limiting level, insertion losses, bandwidth) for different spin-wave modes and transducer lengths. 
An analytical theory has been formulated to describe the fundamental physical processes, and a numerical model has been developed to quantitatively describe the insertion losses and power characteristics of the FSLs.
Additionally, the perspective of the spin-wave devices is discussed, including the possibility of simultaneously integrating three devices into one: a frequency-selective limiter, an RF filter, and a delay line, allowing for more efficient use of space and energy.
\end{abstract}


\maketitle

\section{I. Motivation} \label{sec:Introduction}
The ongoing demand for faster and more efficient 5G communication systems requires all radio-frequency (RF) devices to adapt from the standard frequency ranges of 694–790~MHz (EU 5G low-band) and 3.4–3.8~GHz (EU 5G mid-band) to higher operating frequencies of 24.25–27.5~GHz (EU 5G high-band).

Frequency Selective Limiters (FSLs) are essential devices for protecting electronics from large input signals that can cause overload and potential damage. Historically, ferrite-based FSLs were used for power limiting, but due to their bulkiness, they were replaced by semiconductor-based power limiters \cite{PIN1, Giarola}, which enabled device miniaturization and on-chip integration. However, semiconductor-based power limiters exhibit significant electrical noise and switching delays when operating in the GHz range \cite{Krowne}, have complex designs, and typically attenuate all signals equally, regardless of their magnitude \cite{Adam1}.

In contrast, ferrite spin-wave-based power limiters maintain their efficiency at high GHz ranges; they are passive, low-cost, simple to design, and provide frequency-selective attenuation of incoming signals \cite{Adam2}, giving them their name, frequency selective limiters. This means that a large signal at one frequency undergoes attenuation while a small signal at a different frequency remains unattenuated, resulting in an improved signal-to-noise ratio. This is particularly relevant for information and communication technologies, such as Mobile User Objective Systems (MUOS) \cite{Adam3}, Global Positioning Systems (GPS) \cite{Adam3, Lin, Shukla}, or communication links in self-driving vehicles \cite{Shukla}. For these systems, it is crucial to suppress parasitic signals, which may be either low-power or high-power, while simultaneously detecting a low-power signal of interest \cite{Adam3}.

Spin-wave (SW)-based FSLs are particularly promising in this context, but studies to date have been limited to macroscopic sizes (centimeter to millimeter size) and frequency ranges below 15 GHz -- see the recent review \cite{Khrysreview} focused on SW RF application covering the classical and the most recent results, and discussing SW filters, channelizers, phase shifters, delay lines, FSLs, signal-to-noise enhancers, resonators, and directional couplers. Therefore, the miniaturization of FSLs to nanoscale and investigation of their performance over a broader frequency range is of interest.

In this work, we report on the nanoscale frequency selective limiters. The proposed FSLs are based on spin-wave transmission between excitation and detection transducers fabricated on a Yttrium Iron Garnet thin film affected by multi-magnon scattering when SW amplitude reaches a certain threshold value.
The FSLs were tested across a broad frequency range, including $4\,\mathrm{GHz}$ (EU mid-band), $9\,\mathrm{GHz}$, and $25\,\mathrm{GHz}$ (EU high-band), for two fundamental spin-wave modes: Damon-Eshbach (DE, with the applied field in-plane and perpendicular to SW propagation) and Backward Volume (BV, with the applied field in-plane and parallel to spin-wave propagation). The power characteristics and insertion losses for different SW modes and transducer lengths were extracted from the measurements. The power characteristic curves exhibit saturation at a power-limiting level of 4~GHz, and they exhibit 5~dB oscillation around the power-limiting level values of 9~GHz and 25~GHz. Numerical and analytical models were developed to get deeper insight into the physical processes and to obtain quantitative parameters of the devices.

\section{II. The State of the Art} \label{sec:STAT}
\begin{figure}[hbt]
\centering
   \includegraphics[width=0.48\textwidth]{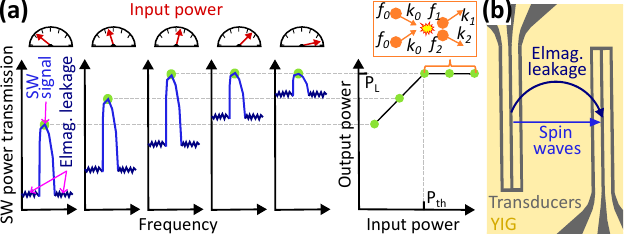}
   \label{fig0}
\caption{(a) (left) Schematic illustration of spin-wave (SW) power transmission for increasing input powers from left to right. The SW signal (light blue) and electromagnetic leakage (dark blue) are indicated for the first plot. The SW signal maximum is highlighted in green. (right) Power limiting characteristics extracted from the SW signal maxima at different input powers. The power threshold $P_{\mathrm{th}}$ and power limiting level $P_{\mathrm{L}}$ are indicated. The multi-magnon scattering, which takes place above the power threshold, is depicted in the orange box. (b) Schema of the spin wave and electromagnetic leakage transmission between the transducers placed on the YIG film.  
}
\end{figure}

The purpose of the power limiter is to maintain output power at the (maximal allowed) power limiting level $P_\mathrm{L}$ if the input power overcomes a certain power threshold level $P_\mathrm{th}$, as shown in Fig.~\hyperref[fig0]{1(a)}. There are two types of ferrite-based FSLs based on either spin-wave transmission or absorption.

For the FSLs based on spin-wave absorption \cite{Khrysreview}, the limiting device consists either of a sphere resonator \cite{Spencer, Okwit, Elliot1, Elliot2} or a conductor placed on the ferrite film \cite{Stitzer1, Stitzer2, Adam4, Yang, Lin}. In this case, the power limiting effect is based on parametric spin-wave generation at half of the frequency $f/2$, which could be direct photon $\to$ (magnon + magnon) three-wave splitting, or three-magnon splitting of directly excited magnon at $f$ (photon at $f$ $\to$ magnon at $f$ $\to$ (magnon + magnon) at $f/2$), or both, depending on the geometry and bias field orientation \cite{3magnon}. If the ferrite material is exposed to a sufficiently large RF power, this power is partially driven away by the excitation of the secondary spin waves. This approach has the advantage of small insertion losses since the energy is primarily kept within the electromagnetic signal domain \cite{Filter1, Filter2, Filter3, Filter4}.

For the FSLs based on spin-wave transmission \cite{Khrysreview}, the limiting device consists of a pair of inductive transducers (analog of interdigital transducers (IDTs) for Surface Acoustic Waves) placed on a ferrite film \cite{Adam1, Shukla} to convert the microwave signal to a spin-wave signal and vice versa; see Fig.~\hyperref[fig0]{1(b)}.
The conversion is due to the interaction of the alternating current in the transducer with the aligned magnetic moments in the magnetic material mediated by dynamic magnetic fields. The magnetic moments alignment is performed by applying an external magnetic field. The spin waves are excited if the frequency of the microwave signal matches the SW frequency defined by the dispersion relation.
Efficient excitation requires the SW wavenumber, determined by the dispersion relation, to match the transducer's geometry and size. The excited spin waves propagate in the magnetic film until they reach the detection (output) transducer, in which the RF voltage is induced by the alternating magnetic field created by the spin waves.  

 In this case, the power limiting effect is based on the multi-magnon scattering \cite{Multimag1, Multimag2}, schematically depicted in the orange inset in Fig.~\hyperref[fig0]{1(a)}. If low input power is sent to the excitation transducer located near the ferrite film, which is exposed to the external magnetic field, spin waves are generated and propagate with low losses towards the detection transducer as no multi-magnon scattering is present. When the input power reaches a power threshold, some of the original magnons generated by the excitation transducer are scattered into new magnons of different frequencies and wave vectors; thus, they can not reach the detection transducer and do not contribute to the transmitted signal. During the scattering processes, both energy and momentum are conserved. Above the threshold, the scattering probability and, thus, the power dissipation increase,  providing approximately constant output power for input powers above the power threshold; see Fig.~\hyperref[fig0]{1(a)}.

Another physical mechanism that might contribute to the experimental findings is the nonlinear drop in the excitation efficiency of the spin waves with the increasing applied RF power. This drop can be due to the shift of the spin-wave dispersion locally under the antenna region, which is caused by the change in the effective magnetization \cite{Qi}.

In this study, we focused on the FSLs based on SW transmission, as it has better frequency selectivity and lower power threshold \cite{Shukla, Adam1, Adam5} compared to the FSLs based on SW absorption. Achieving the lowest possible power threshold is crucial for GPS applications, as the powers of the receiving signals are $-130\,\mathrm{dBm}$ (signal of interest) and $-60\,\mathrm{dBm}$ (disruptive signal) \cite{Adam3}. 

To reach such a low power threshold, reducing the thickness of the magnetic film is essential, as predicted by Adam and Winter \cite{Adam5}. The material of choice in their research and till now is ferrite isolator Yttrium Iron Garnet (YIG) due to its smallest SW damping \cite{YIGsaga, YIGmagnonics, Lastsaga}. The smallest damping also leads to the narrowest linewidth, resulting in the best frequency selectivity, determining how close in frequency small signals can be to larger ones without being attenuated. Over the last decade, the methodology for the growth of nm-thick high-quality YIG was developed using Liquid Phase Epitaxy (LPE) \cite{Carsten, Jouseff}, Pulsed Laser Deposition (PLD) \cite{Caroline, Anane} and sputtering \cite{Schmidt, YIGWu} opening the access to nano-scale SW RF applications. The power threshold can also be lowered by reducing the conductor size and shape \cite{Stern}, as the power threshold is determined by the critical magnitude of the dynamic field component.
Narrowing the conductor increases the current density within it; consequently, for the same input power, the dynamic field component around the conductor is higher, leading to a lower power threshold.

\section{III. Sample Fabrication and Experimental Setup} \label{sec:sample}
In our experiments, we used $97\,\mathrm{nm}$ thin LPE YIG grown on $500\,\upmu\mathrm{m}$ thick (111) Gallium Gadolinium Garnet (GGG) substrate. We fabricated microwave Coplanar Waveguide (CPW) transducers on the YIG film using e-beam lithography and metal evaporation technique. 

\begin{figure}[hbt]
\centering
   \includegraphics[width=0.48\textwidth]{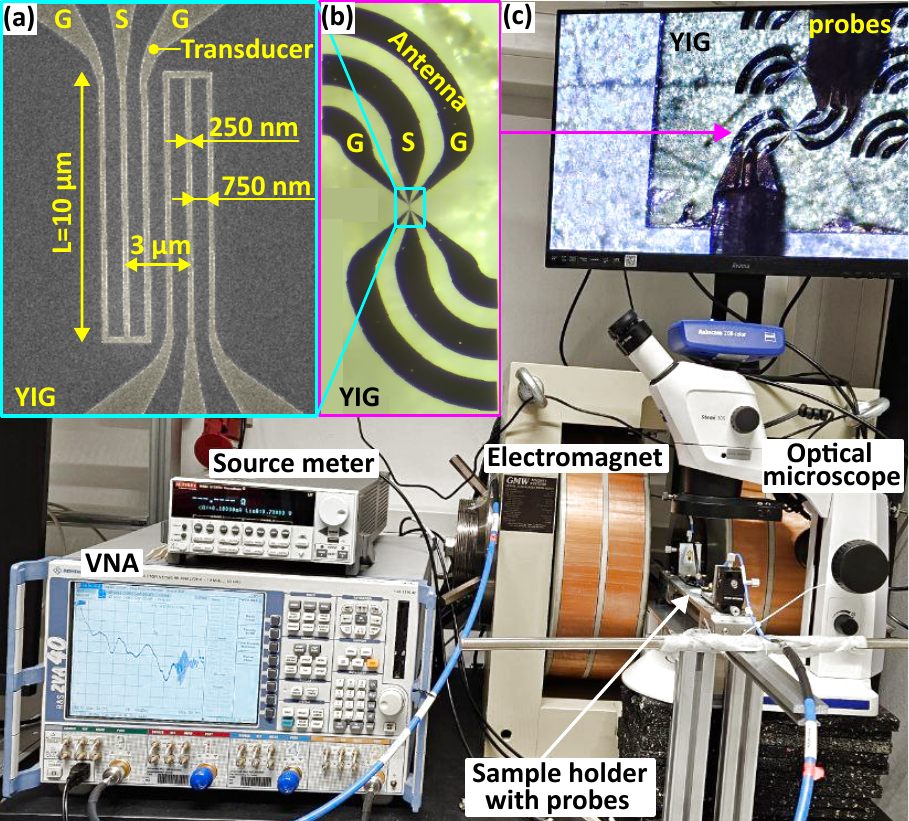}
   \label{fig1}
\caption{Fabricated structures and experimental setup. (a) SEM image of fabricated  $10\,\upmu\mathrm{m}$ long CPW transducers of a center-to-center distance of $3\,\upmu\mathrm{m}$ on $97\,\mathrm{nm}$ thin YIG film. The width of the signal (S) and ground (G) conductors is $250\,\mathrm{nm}$, and the spacing between the conductors is $750\,\mathrm{nm}$. 
(b) Image from the optical microscope of the whole structure together with the curved transducer pads needed for contacting the picoprobes.
(c) CPW transducers are connected via picoprobes and coaxial cables to VNA. 
}
\end{figure}

First, the adhesion layer (AR 300-80), positive PMMA resist (AR-P 672.08), and conductive layer (Electra PC 50-90) were spin-coated and baked on a clean YIG surface. Then, the microwave transducer structures were patterned using an e-beam writer (acc. voltage 30~kV, beam current 500~pA). Afterward, the conductive layer was dissolved, and the resist developed. To remove the resist residuals and adhesion layer from the developed areas, 15 seconds of oxygen plasma etching was applied. In the next step, $10\,\mathrm{nm}$ of titanium (adhesion layer), $320\,\mathrm{nm}$ of copper, and $20\,\mathrm{nm}$ of gold (capping layer) was deposited on the sample using electron-beam physical vapor deposition. As the last step, the lift-off was performed. We fabricated CPW transducers with the lengths $L$ of $10\,\upmu\mathrm{m}$ and $100\,\upmu\mathrm{m}$, and center-to-center distance between the signal lines of $3\,\upmu\mathrm{m}$. Both the signal and ground lines have a width of $250\,\mathrm{nm}$, and the spacing between them is $750\,\mathrm{nm}$; see Fig.~\hyperref[fig1]{2(a)}.

The measurement of SW transmission as a function of frequency for various input power levels, as shown in Fig.~\hyperref[fig0]{1(a)}, is typically performed using a Vector Network Analyzer (VNA). The VNA is well-suited for this application as it is capable of both generating and detecting microwave signals, and this technique is called Propagating Spin-Wave Spectroscopy (PSWS) \cite{Vlaminck, vanatka, milikelvin, csaba}.

Our experimental setup consists of VNA (R$\&$S, ZVA~40), which is connected via high-frequency coaxial cables to picoprobes (GGB, 40A) placed on the microwave transducers, see Fig.~\hyperref[fig1]{2(c)}. The sample holder with the microwave transducers is located between the poles of an electromagnet. An optical microscope is used for contacting the picoprobes to the transducer pads, which are curved and thus allow for contacting the sample for both SW modes; see Fig.~\hyperref[fig1]{2(b)}. A sourcemeter is used to check a proper connection between picoprobes and transducers.

\begin{figure*}[hbt!]
\centering
\includegraphics[width=\textwidth]{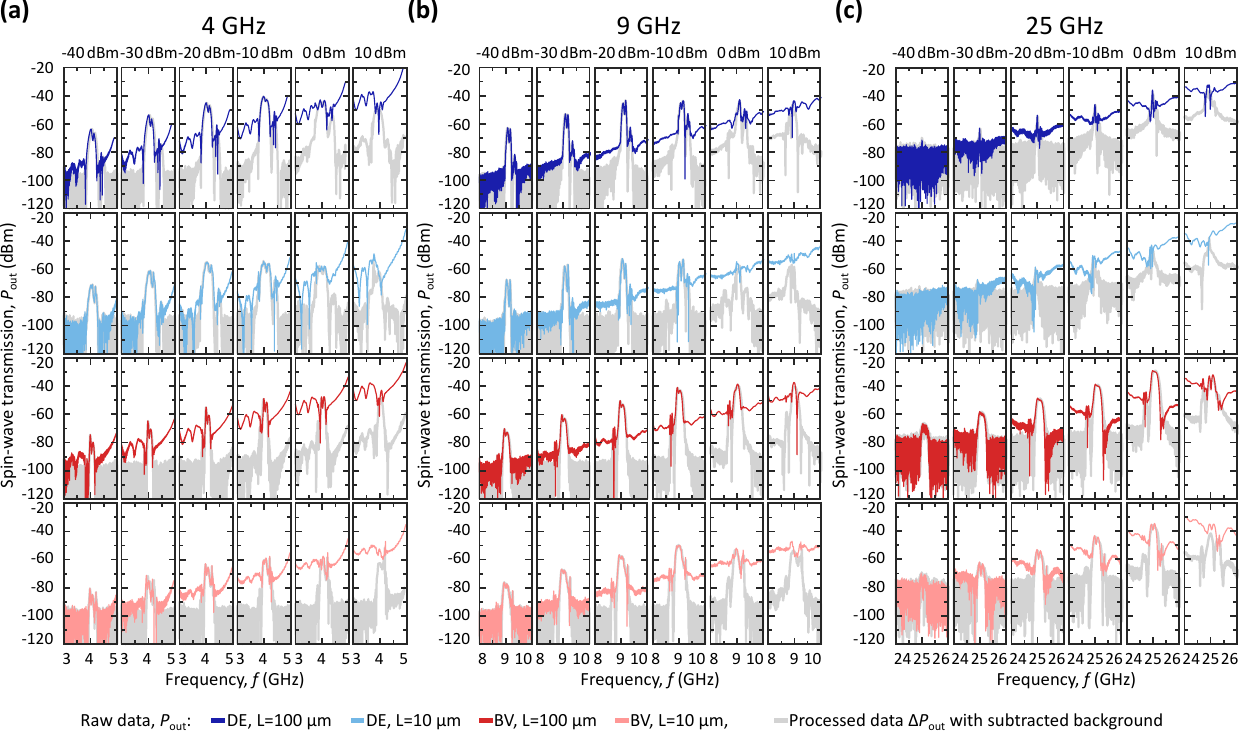}
\label{fig2}
\caption{{Measured spin-wave power transmission using $100\,\upmu\mathrm{m}$ and $10\,\upmu\mathrm{m}$ long CPW transducers for Damon-Eshbach (DE, in blue) and Backward Volume (BV, in red) modes at frequency ranges around (a) $4\,\mathrm{GHz}$, (b) $9\,\mathrm{GHz}$ and (c) $25\,\mathrm{GHz}$ for input powers from $-40\,\mathrm{dBm}$ up to $10\,\mathrm{dBm}$.} Raw data (in red and blue) are plotted together with the processed data with subtracted background (in grey) as a guideline for peak position. Raw data are plotted in dark [light] color (blue for DE or red for BV) when measured using $100\,\upmu\mathrm{m}$ [$10\,\upmu\mathrm{m}$] long transducers. The applied magnetic fields are depicted in Table~\ref{tab2}. The data shown in this figure are also in Supplementary Material \cite{SupplMat} at a larger scale and with a grid, and the measured data are available at \cite{Phaidra}.
}
\end{figure*}

\section{IV. Results} \label{sec:results}
We performed systematical measurements using the fabricated transducers described in the previous section. The experimental data used in this study are available in the Phaidra repository \cite{Phaidra}. The spin-wave transmission was investigated for the input powers $P_\mathrm{in}$ in the range from $-40\,\mathrm{dBm}$ to $10\,\mathrm{dBm}$ with the step of $1\,\mathrm{dBm}$ in the frequency ranges around $4\,\mathrm{GHz}$, $9\,\mathrm{GHz}$, and $25\,\mathrm{GHz}$ with the frequency span of $2\,\mathrm{GHz}$ for DE and BV modes. VNA parameters used for the measurements were bandwidth $1\,\mathrm{kHz}$, frequency step $200\,\mathrm{kHz}$, and no averaging.
The measured complex SW transmission $S_{21}$ was recalculated into the output power $P_\mathrm{out}$ as follows
\begin{equation}
    P_\mathrm{out} \textcolor{gray}{(\mathrm{dBm})}  = 20\mathrm{log}_{10}(|S_{21}|) \textcolor{gray}{(\mathrm{dB})} + P_\mathrm{in} \textcolor{gray}{(\mathrm{dBm})},
\end{equation}
where $S_{21}=S_{21, \mathrm{Re}}+i S_{21, \mathrm{Im}}$ is reconstructed from real $S_{21, \mathrm{Re}}$ and imaginary $S_{21, \mathrm{Im}}$ S-parameters measured by VNA.
For a better understanding of the measured signal, the processed data $\Delta P_\mathrm{out}$ with background subtraction are calculated using the formula
\begin{equation}
    \Delta P_\mathrm{out} \textcolor{gray}{(\mathrm{dBm})}  = 20\mathrm{log}_{10}(|S_{21}-S_{21, \mathrm{REF}}|) \textcolor{gray}{(\mathrm{dB})} + P_\mathrm{in} \textcolor{gray}{(\mathrm{dBm})},
\end{equation}
where $S_{21, \mathrm{REF}}$ is the reference signal measured at different magnetic fields (detuned by $100\,\mathrm{mT}$) with no spin-wave signal. This reference signal is the electromagnetic leakage between the transducers (cross-talk); see Fig.~\hyperref[fig0]{1(b)}. In our case, the electromagnetic leakage is mostly due to the transducer pads and can be significantly reduced by their proper design. The data with the background subtraction thus shows only the spin wave signal with no electromagnetic leakage.

The SW power transmission spectra measured using $10\,\upmu\mathrm{m}$ and $100\,\upmu\mathrm{m}$ long CPW transducers are shown in Fig.~\ref{fig2}. The peak in the measured data is the spin-wave signal, and its background is the electromagnetic leakage, similarly as depicted in Fig.~\hyperref[fig0]{1(a)}. The measured raw data are plotted alongside the processed data with a subtracted background (in grey) to serve as a guideline for peak position visibility. The applied in-plane magnetic fields are depicted in Table~\ref{tab2} for both SW modes and all frequency ranges.

\begin{figure*}[hbt!]
\centering
   \includegraphics[width=\textwidth]{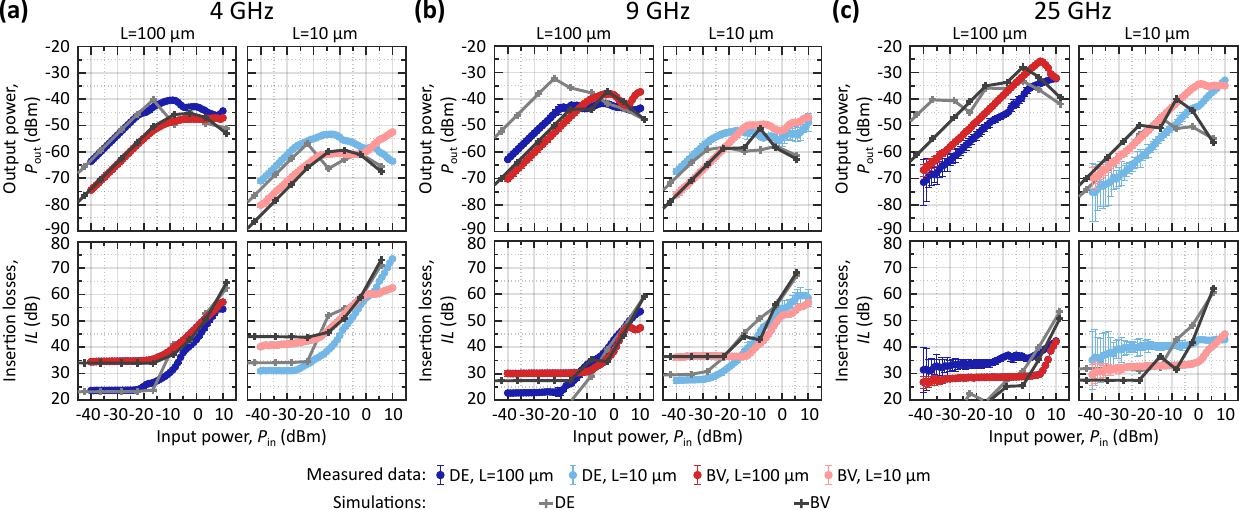}
   \label{fig3}
    \caption{Power characteristic and insertion losses extracted from raw data at the peak maximum at (a) $4\,\mathrm{GHz}$, (b) $9\,\mathrm{GHz}$ and (c) $25\,\mathrm{GHz}$ measured using $100\,\upmu\mathrm{m}$ (dark color) and $10\,\upmu\mathrm{m}$ (light color) long CPW transducers for Damon-Eshbach (DE, in blue) and Backward Volume (BV, in red) modes. The simulated trends are plotted in light grey for DE and in dark grey for BV. The applied magnetic fields are depicted in Table~\ref{tab2}.}
\end{figure*}

From the raw data of the measured spin wave power transmission in Fig.~\ref{fig2} (this figure is available in the Supplementary Material \cite{SupplMat} at a larger scale and with a grid.), we can see increasing electromagnetic leakage (the background around the peak) with increasing input power, as expected. On the other hand, the spin wave signal (the peak) increases for the small input powers, and then it becomes constant for the higher input powers due to the energy dissipation into multi-magnon scattering. 

To see the power characteristic better, the output power values ($P_\mathrm{out}$, from raw data) were extracted at the peak maximum from Fig.~\ref{fig2}; and they are depicted in Fig.~\ref{fig3} for both DE and BV modes at all investigated frequency ranges. Note that the maximum of the SW transmission was not always precisely at 4~GHz, 9~GHz, or 25~GHz but may be shifted by $\pm70\,\mathrm{MHz}$.
Below a certain power threshold, the extracted power characteristics increase linearly; thus, the output power is proportional to the input power. As the power increases, a threshold is reached, and the transmitted power no longer increases with increasing input power due to the nonlinear four-magnon scattering process \cite{Gurevich-Melkov, LvovBook}. 

In the four-magnon scattering process, a pair of primary magnons excited by the input transducer scatter into a pair of secondary magnons. The wavevectors, frequencies, and phases of the secondary magnons are not strictly defined, but the process requires the conservation of momentum and energy. Often, the secondary magnons propagate in a different direction than the primary ones (i.e., have a large wave vector component perpendicular to that of the primary magnons) and/or have much shorter wavelengths, being in the dipolar-exchange region of the SW dispersion relation. Therefore, they cannot be detected or even reached by the output transducer \cite{MagnonTransistor}. As a result, the SW signal transmitted between the transducers is reduced because part of its energy is transferred to the secondary magnons and lost within the magnetic film. The opposite case, when secondary magnons are close in $\bm{k}$-space with the primary ones, is also possible; it is called the modulation instability \cite{LvovBook} and should be avoided when designing FSLs. 

Some of the extracted curves in Fig.~\ref{fig3} exhibit the ideal trend - the transmitted power for the input powers above the power threshold is of a constant value, precisely determining the power limiting level. In some other cases, there are variations in the power-limiting level value around 5 dB. These variations are due to complex nonlinear physics and are defined by the interaction strength between the primary and secondary magnon groups, the competition between different magnon groups, and the exact mechanism responsible for the saturation of the parametric process \cite{LvovBook, Andrii}. However, from an applied point of view, it is important that no transmission of the large-magnitude signal through the device is possible, and thus, the FSL protects the device.   

The insertion losses $IL$, one of the key parameters for all microwave devices, were also extracted at the peak maximum from Fig.~\ref{fig2}. They are recalculated from the complex SW transmission as follows
\begin{equation}
    IL \textcolor{gray}{(\mathrm{dB})}  = |20\mathrm{log}_{10}(|S_{21}|)| \textcolor{gray}{(\mathrm{dB})}.
\end{equation}
The relation between the insertion losses, input, and output power is
\begin{equation}
    P_{\mathrm{out}} \textcolor{gray}{(\mathrm{dBm})}  = P_{\mathrm{in}} \textcolor{gray}{(\mathrm{dBm})} - IL \textcolor{gray}{(\mathrm{dB})},
    \label{ILeq}
\end{equation}
therefore, in the ideal case, the insertion losses should be close to $0\,\mathrm{dB}$ for the input powers below the power threshold (linear region). In our case, the insertion losses are above $20\,\mathrm{dB}$, but they can be significantly reduced (ultimately down to a few dB \cite{IL1, IL2, IL3, connely, Budapest, Florian}) by optimizing the transducer design and the YIG film thickness.  

Our findings depicted in Fig.~\ref{fig3} clearly demonstrate that nanoscale FSLs work well in a broad frequency range. To compare the power characteristics conveniently, the power threshold $P_\mathrm{th}$ and power limiting level $P_\mathrm{L}$ were extracted. The power threshold is the input power value at which the power limiting starts and the power limiting level is the value of the constant output power; these characteristics are indicated in Fig.~\hyperref[fig0]{1(a)}.

\begin{figure}[hbt]
\centering
   \includegraphics[width=0.48\textwidth]{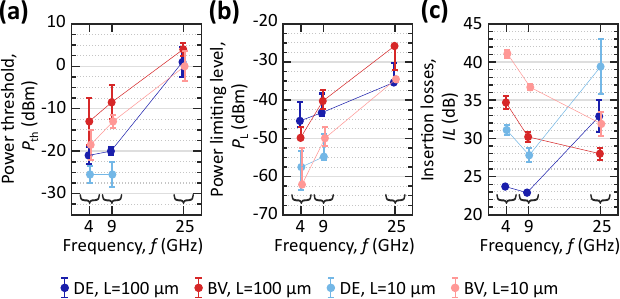}
\label{fig4}
\caption{Extracted (a) power threshold, (b) power limiting level and (c) insertion losses from data depicted in Fig.~\ref{fig3} for Damon-Eshbach (DE, in blue) and Backward Volume (BV, in red) for CPW transducers of the length of $10\,\upmu\mathrm{m}$ (in light color) and $100\,\upmu\mathrm{m}$ (in dark color) at $4\,\mathrm{GHz}$, $9\,\mathrm{GHz}$ and $25\,\mathrm{GHz}$. The insertion losses were extracted in the linear region at the input power of $-30\,\mathrm{dB}$. For better visibility, there is a slight offset in frequency between the extracted values measured using different transducer lengths and spin-wave modes at the same frequency.
}
\end{figure}

The extracted values of power threshold, power limiting level, and insertion losses (extracted in the linear region for the input power of $-30\,\mathrm{dBm}$) are shown in Fig.~\ref{fig4}. The values were extracted for both SW modes and all frequency ranges, except the DE mode at $25\,\mathrm{GHz}$ measured using $10\,\upmu\mathrm{m}$ long transducers, as it does not exhibit a clear power threshold and limiting level. 
For better visibility, the data are a bit shifted along the x-axis (in frequency) for different transducer lengths and SW modes measured at the same frequency. The values of power threshold, power limiting level, and insertion losses are also depicted in Table~\ref{tab2} for $100\,\upmu\mathrm{m}$ long CPW for both SW modes. In the following list, the main findings are summarized.

\begin{table}[hbt!]
\centering
\caption{The applied magnetic field $B$, calculated group velocities $v_g$, decay lengths $\delta$, extracted power threshold $P_{\mathrm{th}}$, power limiting level $P_{\mathrm{L}}$, insertion losses $IL$, bandwidths $BW$ and calculated delay time $T$ for Damon-Eshbach (DE) and Backward Volume (BV) modes at $4\,\mathrm{GHz}$, $9\,\mathrm{GHz}$ and $25\,\mathrm{GHz}$. 
Group velocities and decay lengths were calculated at $k=3\,\mathrm{rad/\upmu m}$ using the Kalinikos and Slavin model \cite{kalinikos}. 
The values for insertion losses were extracted at the input power $-30\,\mathrm{dBm}$ and for the bandwidth at $-20\,\mathrm{dBm}$. The power threshold, power limiting level, and insertion losses are listed for both $100\,\upmu\mathrm{m}$,  $10\,\upmu\mathrm{m}$ and long CPW transducers used for SW transmission.
The delay time, $T$, was calculated by dividing the transducer's center-to-center distance of $3\,\upmu\mathrm{m}$ by the spin-wave group velocity.}
\label{tab2}
\resizebox{\columnwidth}{!}{%
\begin{tabular}{ccllcccccc}
\hline\hline
\multicolumn{4}{c}{\multirow{2}{*}{}} & \multicolumn{2}{c}{$4\,\mathrm{GHz}$} & \multicolumn{2}{c}{$9\,\mathrm{GHz}$} & \multicolumn{2}{c}{$25\,\mathrm{GHz}$} \\
\multicolumn{4}{c}{} & DE & BV & DE & BV & DE & BV \\
\cmidrule(lr){5-6} \cmidrule(lr){7-8} \cmidrule(lr){9-10}
\multicolumn{4}{c}{$B\,\mathrm{(mT)}$} & 71 & 87 & 244 & 255 & 807 & 823 \\
\multicolumn{4}{c}{$v_g\,\mathrm{(m/s)}$} & 606 & 312 & 302 & 434 & 144 & 512 \\
\multicolumn{4}{c}{$\delta\,\mathrm{(}\upmu\mathrm{m)}$} & 107.8 & 53.8 & 25.6 & 37.1 & 4.6 & 16.5 \\
\hline
\parbox[t]{2mm}{\multirow{3}{*}{\rotatebox[origin=c]{90}{$100\,\upmu\mathrm{m}$}}} & \multicolumn{3}{c}{$P_{\mathrm{th}}\,\mathrm{(dBm)}$} & -21.0$\pm$2.0 & -13.0$\pm$5.5 & -20.0$\pm$1.0 & -8.5$\pm$4.0 & 1.0$\pm$3.5 & 4$\pm$1.5 \\
 & \multicolumn{3}{c}{$P_{\mathrm{L}}\,\mathrm{(dBm)}$} & -45.4$\pm$3.3 & -49.8$\pm$1.4 & -43.3$\pm$1.3 & -40.2$\pm$2.5 & -35.3$\pm$1.5 & -25.8$\pm$3.1 \\
 & \multicolumn{3}{c}{$IL\,\mathrm{(dB)}$} & 23.7$\pm$0.1 & 34.7$\pm$0.8 & 22.9$\pm$0.2 & 30.2$\pm$0.7 & 32.9$\pm$2.1 & 28.0$\pm$0.8 \\
\hline
\parbox[t]{2mm}{\multirow{3}{*}{\rotatebox[origin=c]{90}{$10\,\upmu\mathrm{m}$}}} & \multicolumn{3}{c}{$P_{\mathrm{th}}\,\mathrm{(dBm)}$} & -25.5$\pm$2.0 & -18.5$\pm$3.5 & -25.5$\pm$3.3 & -13.0$\pm$1.5 & -- & 0$\pm$3.5 \\
 & \multicolumn{3}{c}{$P_{\mathrm{L}}\,\mathrm{(dBm)}$} & -57.5$\pm$5.1 & -62.1$\pm$4.8 & -54.8$\pm$3.5 & -50.0$\pm$2.5 & -- & -34.5$\pm$0.2 \\
 & \multicolumn{3}{c}{$IL\,\mathrm{(dB)}$} & 31.1$\pm$0.7 & 41.1$\pm$0.5 & 27.8$\pm$0.9 & 36.7$\pm$0.4 & 39.4$\pm$3.6 & 31.9$\pm$1.6 \\
\hline
\multicolumn{4}{c}{$BW\,\mathrm{(MHz)}$} & 280$\pm$32 & 160$\pm$53 & 170$\pm$49 & 180$\pm$20 & 50$\pm$29 & 220$\pm$5 \\
\multicolumn{4}{c}{$T\,\mathrm{(ns)}$} & 4.95 & 9.61 & 9.93 & 6.91 & 20.83 & 5.86 \\

\hline\hline
\end{tabular}%
}
\end{table}

\vspace{0.15cm}
\textbf{\textit{The comparison of different spin-wave modes}}

\begin{itemize}
    \item \textbf{The power threshold} $P_{\mathrm{th}}$ is always smaller for DE than for BV (see Fig.~\hyperref[fig4]{5(a)} and Table~\ref{tab2}) because of the better transducer excitation efficiency in DE mode \cite{DEvsBV}, since two (in-plane and out-of-plane) rather than only one component of the dynamic magnetic field contribute to the spin-wave excitation. This results in reaching the critical magnitude of the dynamic field component and power threshold at lower input powers.

    \item \textbf{The power limiting level} $P_{\mathrm{L}}$ is higher for DE in the low-frequency range and for BV in the high-frequency range. This dependence is mainly caused by SW group velocity and is explained in the next section. 
    
    \item \textbf{Low-power insertion losses} $IL$ are smaller for DE waves at lower frequencies and for BV at high frequencies. The first underlying reason is the above-mentioned transducer efficiency, in particular, the nonreciprocity of DE SWs excitation by a microstrip transducer \cite{Schneider_PRB2008}. At higher frequencies, however, increased propagation losses of DE SWs due to lowering the group velocity overcomes the excitation efficiency gain, and BV SWs show smaller $IL$.
    
\end{itemize}

\textbf{\textit{The comparison of different transducer lengths}}
\begin{itemize}
    \item \textbf{The power threshold} $P_{\mathrm{th}}$ is higher for longer transducers because longer transducers have higher resistance compared to short ones, resulting in a smaller current density for the same applied microwave power. Combined with a higher impedance mismatch, this leads to reaching the critical magnitude of the dynamic field component and power threshold at higher input powers.

    \item \textbf{The power limiting level} $P_{\mathrm{L}}$ is higher for longer transducers because they cover a bigger magnetic volume, leading to a larger transmitted SW signal. The difference is close to 10~dBm, i.e. 10 times, exactly as the difference between the transducer's length; see Fig.~\hyperref[fig4]{5(b)} and Table~\ref{tab2}. 
    The same reason causes a difference in the insertion losses at low power; see Fig.~\hyperref[fig4]{5(c).}
\end{itemize}

\textbf{\textit{Sub-conclusion}}
\begin{itemize}

    \item \textbf{To achieve the lowest possible power threshold}, it is optimal to operate in the DE mode, at least in the studied frequency range. Additionally, to further reduce the power threshold, narrow and thin microwave transducers should be used to ensure strong dynamic magnetic field confinement within the magnetic sample. These transducers should also be short in length to minimize DC resistance. However, it is crucial to balance the trade-off between reducing the transducer size and maintaining impedance matching, as impedance mismatch would reduce power transmission efficiency and result in a higher power threshold.
    
    \item \textbf{To achieve the highest power limiting level and lowest insertion losses}, it is advantageous to operate in the DE mode at lower frequencies and switch to the BV mode at higher frequencies. Additionally, using longer transducers that cover a larger magnetic volume is beneficial; see Fig.~\hyperref[fig4]{5(c)}.
\end{itemize}

\begin{figure*}[hbt!]
\centering
   \includegraphics[width=\textwidth]{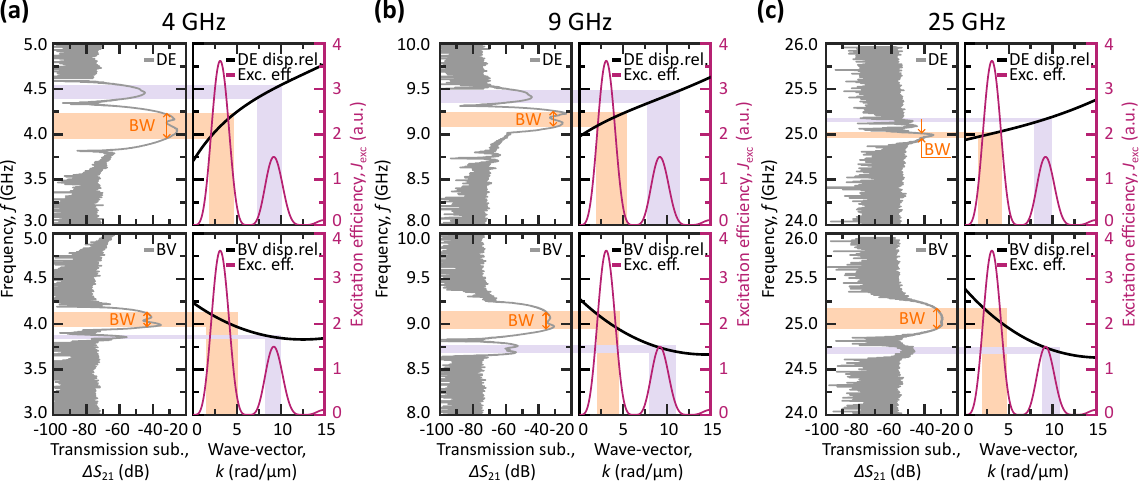}
\label{fig5}
\caption{(left panels) Spin wave transmission with background subtraction (in grey) measured using $100\,\upmu\mathrm{m}$ long CPW transducers at the input power of $-20\,\mathrm{dBm}$ at (a) $4\,\mathrm{GHz}$, (b) $9\,\mathrm{GHz}$ and (c) $25\,\mathrm{GHz}$ for Damon-Eshbach (DE, upper panels) and Backward Volume (BV, lower panels) modes. The Bandwidth $BW$ of the main signal peak defined as $-3\,\mathrm{dB}$ decrease from the average peak maximum is indicated. (right panels) Calculated transducer excitation efficiency (in magenta) and dispersion relation (in black) using the Slavin and Kalinikos model \cite{kalinikos}. The match between the measured transmission and transducer excitation spectra is highlighted in orange for the main peak and in violet for the side peak.
}
\end{figure*}

\section{V. Theory, Simulations and Analysis} \label{sec:Theory}

\subsection{A. Dispersion relation, group velocity, transducer excitation spectrum, and signal bandwidth}

To have a better understanding of the measured signals for DE and BV at different frequency ranges, the measured spin-wave transmission $\Delta S_{21}$ is compared with the dispersion relation and transducer excitation efficiency $J_\mathrm{exc}$, see Fig.~\ref{fig5}. 
For better visibility of the transmitted signal shape, the spin wave transmission with a subtracted background shown in Fig.~\ref{fig5} is recalculated to the dB scale as follows
\begin{equation}
    \Delta S_{21} \textcolor{gray}{(\mathrm{dB})}  = 20\mathrm{log}_{10}(|S_{21}-S_{21, \mathrm{REF}}|) \textcolor{gray}{(\mathrm{dB})}.
\end{equation}
The dispersion relations (black curves in Fig.~\ref{fig5}) were calculated using the analytical model from Kalinikos and Slavin \cite{kalinikos} with these input values: film thickness $t=97\,\mathrm{nm}$, exchange constant $A_\mathrm{ex}=3.6\,\mathrm{pJ/m}$, gyromagnetic ratio $\gamma/2\pi=28\,\mathrm{GHz/T}$, saturation magnetization: $M_\mathrm{s}=140\,\mathrm{kA/m}$, fundamental mode $n=0$, pinning condition: fully unpinned spins at the surface. Values for the applied magnetic field are written in Table~\ref{tab2}.
From the dispersion relation, we also calculated the group velocity as its derivative $v_\mathrm{g} = \frac{2\pi\partial f}{\partial k}$ and decay length $\delta = v_\mathrm{g} \tau$, where $\tau$ is a SW lifetime. It can be calculated as a field derivative \cite{lifetime} $\tau=\big(\frac{\alpha_G 4 \pi^2 f}{\gamma} \frac{\partial f}{\partial B_\mathrm{ext}}\big)^{-1}$, where $\alpha_G$ is a damping constant (in our case $\alpha_G=0.0002$) and $B_\mathrm{ext}$ is the applied field.
The calculated group velocities and decay length at the signal maximum ($k=3\,\mathrm{rad/\upmu m}$) are summarized in the Table~\ref{tab2}.
The SW group velocities decrease with increasing frequencies for DE mode and increase for BV mode. This is in good agreement with measured power limiting level and insertion losses (see Fig.~\hyperref[fig4]{5(b,c)}), which are lower for DE at smaller frequencies and for BV at higher frequencies. Thus, the use of the BV mode seems to be preferable at a high-frequency range.

The transducer excitation efficiency was calculated as a Fourier transform of the normalized current density in the CPW transducer \cite{Vlaminck}, specifically in the in-plane direction and perpendicular to the longitudinal axes of the CPW antennas 
\begin{equation}\label{e:Jk}
   J_\mathrm{exc}=\Bigg|{\frac{1}{ka}\bigg[\mathrm{sin}\Big(\frac{ka}{2}+kb\Big)-\mathrm{sin}\Big(\frac{3ka}{2}+kb\Big)\bigg]+ \frac{\mathrm{sin}\big(\frac{ka}{2}\big)}{\frac{ka}{2}}}\Bigg|^2,
\end{equation}
where $a$ is the conductor width, $b$ is the gap between the signal and ground conductors, and $k$ is the SW wave number. This simple model can describe the excitation efficiency only qualitatively as it assumes an infinitely thin transducer with normalized current density, which is the same everywhere in all conductors. The calculated excitation efficiency using $a=250\,\mathrm{nm}$ and $b=750\,\mathrm{nm}$ as the input values is plotted in Fig.~\ref{fig5} (magenta curves).
The transducer excitation spectrum has first peaks at $k=3\,\mathrm{rad/\upmu m}$ (highlighted in orange) and a second peak at $k=9.2\,\mathrm{rad/\upmu m}$ (highlighted in violet). 
These peaks correspond well to the measured SW transmission peaks in all investigated frequency ranges for both DE and BV modes. The small displacement in the excitation spectra is related to the assumptions of the transducer's finite thickness and the Fourier transform calculation from the current density rather than from the dynamic magnetic field components.

From the first peak of the measured SW transmission, we extracted bandwidth $BW$ as a $3\,\mathrm{dB}$ decrease from the average peak maximum. The extracted values of the bandwidths are written in Table~\ref{tab2}. The extracted bandwidth values follow the same trend as a group velocity - bandwidth decreases for DE and increases for BV with increasing frequency. 

\subsection{B. Analytical description of the instability threshold}

Next, we analyze the nonlinear characteristics of FSL. For this purpose, we performed analytical calculations of the four-magnon instability threshold power $P_{\mathrm{th}}^{\mathrm{SW}}$, which is related to the limiting SW power. We use the classic SW Hamiltonian formalism in the ``effective spin-wave tensor'' notations \cite{Krivosik_PRB2010} assuming uniform dynamic magnetization distribution across the film thickness and, thus, ignoring possible scattering in higher-order thickness modes. This assumption is valid for thin films and SW wave numbers $kt \ll 1$. In terms of Ref.~\onlinecite{Krivosik_PRB2010}, the four-magnon instability we are interested in is described by the Hamiltonian term $\mathcal H = \tilde W_{00,12}a_0^*a_0^*a_1a_2$, where $a_i$ is canonical spin-wave amplitude and $\tilde W_{00,12}$ is the four-magnon interaction coefficient (accounting for possible contribution from nonresonant three-magnon processes, see details in \cite{Krivosik_PRB2010}). This term describes the splitting of two primary magnons $a_0$ into secondary magnons $a_1$ and $a_2$ under momentum and energy conservation rules $2\vec k_0 = \vec k_1 + \vec k_2$, $2\omega_0 = \omega_1 + \omega_2$. 

For a given frequency and wave vector of primary magnons (we set $|\vec k_0| = 3\,\mathrm{rad}/\upmu\mathrm{m}$ corresponding to the excitation maximum of used CPW transducers) we calculated a set of secondary magnons pairs, which satisfy the energy and momentum conservation rules, and determined the maximum value of the four-magnon interaction efficiency $|\tilde W_{00,12}|$ among this set, as the pair of secondary magnons with the highest interaction efficiency determines the instability threshold $P_{\mathrm{th}}^{\mathrm{SW}}$.

\begin{figure}[hbt]
\centering
   \includegraphics[width=0.48\textwidth]{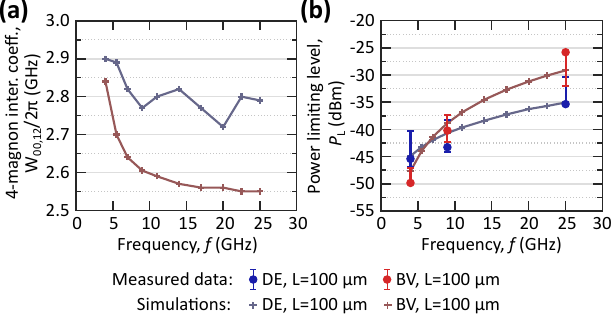}
\label{fig7}
\caption{ (a) Dependence of 4-magnon interaction coefficient $|\tilde W_{00,12}|$ as a function of frequency for Damon-Eshbach (DE, in blue) and Backward Volume (BV, in red) modes. (b) Comparison of calculated SW threshold power $P_{\mathrm{th}}^{\mathrm{SW}}$ (in dark colors) and measured power limiting levels $P_{\mathrm{L}}$ (in bright colors) for DE and BV modes using the $100\,\upmu\mathrm{m}$ long transducers.
}
\end{figure}

The results are presented in Fig.~\hyperref[fig7]{7(a)}. For the BV mode, the interaction efficiency weakly decreases with the frequency of primary magnons. For the DE mode, the dependence is more irregular because of changing the positions of preferential primary magnons in $\vec k$-space and due to nonresonant three-magnon contribution. Nevertheless, overall variation of $|\tilde W_{00,12}|$ does not exceed $10\,\%$. This is an expected feature of second-order parametric (four-magnon) instability, whose efficiency weakly depends on the magnetization precession ellipticity, in contrast to first-order (three-magnon and direct photon $\to$ magnon + magnon) instabilities \cite{Hamadeh2023}.

Threshold amplitude of the primary magnons is given by $|a_{0,\mathrm{th}}|^2 = \sqrt{\Gamma_1 \Gamma_2} / |\tilde W_{00,12}|$, where $\Gamma_i$ is the spin wave damping rate. The corresponding SW threshold power is 
    \begin{equation}\label{e:Pth}
        P_{\mathrm{th}}^{\mathrm{SW}} = \frac{M_s S v_{g} \omega_0 |a_\mathrm{th}|^2}{\gamma} \approx  \frac{\alpha_G M_s S v_{g} \omega_0^2}{\gamma |\tilde W_{00,12}|}\,,
    \end{equation}
where $S = Lt$ is the cross-section of the spin-wave beam, given in our case by the film thickness $t$ and transducer length $L$, $v_{g}$ is the spin-wave group velocity, and the second approximate expression is derived under the assumption that the secondary waves have frequencies close to the primary one. Since the damping rate is proportional to the frequency, $\Gamma \sim \alpha_G \omega$, Eq.~(\ref{e:Pth}) yields almost quadratic increase of the SW threshold power with the operational frequency, $P_{\mathrm{th}}^{\mathrm{SW}} \sim \omega_0^2$, which can be affected by the frequency dependence of the spin-wave group velocity.

Calculated dependencies of the SW threshold power $P_{\mathrm{th}}^{\mathrm{SW}}$ are shown in Fig.~\hyperref[fig7]{7(b)} and are compared to the measured power limiting level. While these quantities are not exactly the same -- the calculations give power in the spin-wave domain $P_{\mathrm{th}}^{\mathrm{SW}}$, while measured power limiting level $P_\mathrm{L}$ is also affected by the output transducer efficiency and propagation losses -- they are closely related, and we find quite a good coincidence of the measurements and calculations. SW threshold power for BV mode increases faster than for DE mode because of the above-mentioned frequency dependence of the group velocity. 

Expression (\ref{e:Pth}) outlines possible ways to control the power limiting level, which becomes particularly challenging at higher frequencies due to its increase, governed by the rule $P_{\mathrm{th}}^{\mathrm{SW}} \sim \omega_0^2$. 
 Lowering group velocity too much is not a good option as it has a consequence of increased propagation losses. Decreasing YIG thickness and length of the transducers (or alternative utilization of finite-width YIG waveguides) are the main options. They, however, require careful transducer design to minimize impedance mismatch. Also, lowering the saturation magnetization, e.g., by YIG doping, could be an option, although one should care about a possible increase of the Gilbert constant of doped YIG. A more involved option could be searching for specific cases of enhanced scattering efficiency (i.e. increased $|\tilde W_{00,12}|$); this option requires deep studies of multimagnon scattering in thin films and possible micro- or nano-structured waveguides. 

\subsection{C. Micromagnetic simulations}
\label{MicromagSim}
Finally, we performed a set of micromagnetic simulations to model the power-limiting effect.
The finite-difference micromagnetic code, {\it magnum.np} \cite{Magnumnp}, was employed to define the transducer geometry and calculate the Oersted field generated by a given current.
This Oersted field contributes to the effective field, enabling the simulation of spin-wave excitation through the Landau-Lifshitz-Gilbert equation
\begin{align}\label{eqn:llg}
\begin{split}
\dot{\vec{m}} &= -\gamma \left[ \vec{m} \times \vec{h}^\text{eff} + \alpha_G \, \vec{m} \times \left( \vec{m} \times \vec{h}^\text{eff} \right) \right], \\
\vec{h}^\text{eff} &= \vec{h}^\text{ex} + \vec{h}^\text{d} + \vec{h}^\text{oe} + \vec{h}^\text{ext},
\end{split}
\end{align}
with the magnetization $\vec{m}$, the effective exchange, dipolar, Oersted and external field contributions $\vec{h}^\text{ex}$, $\vec{h}^\text{d}$, $\vec{h}^\text{oe}$, and $\vec{h}^\text{ext}$, respectively.

The overall efficiency $\eta$ of the transducer can be derived by evaluating the power of the excited spin wave $P^\text{sw}$. It can be decomposed into an Ohmic contribution $\eta^\text{ohm}$ and reflection losses $\eta^\text{refl}$
\begin{align}
  \eta^\text{ohm} &= \frac{P^\text{sw}}{P^\text{sw} + P_0}, \\
  \eta^\text{refl} &= 1 - \Gamma^2, \quad \Gamma \approx \frac{R^\text{sw} + R_0 - \SI{50}{\ohm}}{R^\text{sw} + R_0 + \SI{50}{\ohm}}, \\
  \eta &= \eta^\text{ohm} \; \eta^\text{refl},
\end{align}
with the Ohmic losses in the transducer $P_0 = R_0 I^2/2$, the Ohmic resistance $R_0$, and the total current $I$. The spin-wave resistance $R^\text{sw}$ is derived from the spin-wave power via $P^\text{sw} = R^\text{sw} I^2 / 2$, assuming a common characteristic impedance of external microwave line of $\SI{50}{\ohm}$. The Ohmic resistance is $R_0=19.4\,\Omega$ for $10\,\upmu\mathrm{m}$ long transducers and $R_0=72.2\,\Omega$ for $100\,\upmu\mathrm{m}$ long transducers. This model is somewhat simplified as it ignores the imaginary part of the spin-wave radiation impedance and transducer impedance, as well as possible nonuniform current distribution along the transducers due to spin-wave load.

The power of the excited spin-wave $P^\text{sw}$ is determined by performing a (discrete) time-derivative of the micromagnetic energy
\begin{align}
P^\text{sw} = \frac{\partial E^\text{sw}}{\partial t}, \quad E^\text{sw} = \int\limits_{\Omega_d} e[\vec{m}] \, d\Omega,
\end{align}
with the micromagnetic energy $e[\vec{m}]$. The integration is performed over a limited detection region $\Omega_d$, which is located on one side of the transducer.
This approach suppresses localized excitations that do not result in propagating spin waves, as well as the spin-wave contributions propagating in the wrong direction. 
We also set the damping constant $\alpha_G = 0$ to simulate only the transducer's excitation efficiency. Propagation losses could be easily considered by setting $\alpha_G \ne 0$ and positioning the detection region at the location of the secondary transducer.

Finally, the insertion losses are calculated as the square of the excitation efficiency of the individual transducer
\begin{align}
IL \textcolor{gray}{(\mathrm{dB})} = -10 \, \log_{10}({\eta^2}).
\end{align}

The simulated insertion losses, along with the recalculated output power curves, are shown in grey in Fig. \ref{fig3} (light grey for DE and dark grey for BV). The output power was derived from the insertion losses using equation \ref{ILeq}. Despite the simplicity of the developed numerical model and the complexity of the physical processes observed in the experiment, the numerical results show good quantitative agreement with the experimental data, in particular with respect to the insertion loss in both the linear and nonlinear regime, as well as the power thresholds. In particular, the agreement is particularly robust at a frequency of $4\,\mathrm{GHz}$ and for long transducer lengths of $100\,\upmu\mathrm{m}$. The numerical model has shown good accuracy in both BV and DE configurations. However, the agreement is less precise for short SW transducers, which can be attributed to the experimental assumption of an ``unlimited'' YIG film size, while the simulated structure size is limited to the length of the transducers. As a result, the spin wave energy may propagate at an angle and only partially reach the output antenna in the experiment.

The discrepancies between the experimental and numerical results become more pronounced at higher frequencies, with differences of up to 10 dB in both output power and threshold power. These deviations in output power are due to the aforementioned simplifications inherent in the recalculation of the FSL characteristics from micromagnetic simulations, which cannot be fully resolved by standard micromagnetic software, as well as due to the finite width of the simulated structure, as mentioned above. 

At the same time, the proposed and validated numerical approach has significant potential for designing more complex FSLs where analytical methods are impractical, such as those based on structured films or waveguides. The developed model can be used in future engineering research to optimize transducer design and enhance the performance of spin-wave devices.

\section{VI. Perspective of Spin-Wave Devices} 
\label{Discussion}

Spin-wave RF technology, originally termed Magnetostatic Wave-based technology, was extensively studied a few decades ago \cite{Adam3, Giarola, Khrysreview, oldreview1, oldreview2} but was ultimately surpassed by Surface Acoustic Waves (SAW) and semiconductor technologies.
 However, conventionally used SAW-based filters and delay lines do not allow operating at a high-frequency range, such as the $25\,\mathrm{GHz}$ required for a new generation of 5G telecommunication systems. This restriction arises from the fact that the damping and exposure limitation of interdigitated transducers of SAW-based devices drastically increases with increasing frequency \cite{SAW1, SAW2}.
One possible solution considered nowadays is using Bulk Acoustic Waves (BAW) based filters and delay lines instead, but they suffer from high damping at a high-frequency range and have high requirements for the isolation of the BAW from the substrate, which leads to complicated fabrication \cite{SAW1}.
Therefore, alternative solutions are required to develop filters and delay lines that can operate efficiently and conveniently in the high-frequency range.

Spin waves are promising candidates for RF filters and delay lines that aim to operate at these higher frequencies effectively. It is possible to design SW-based RF filters due to the finite width of the SW band in the SW transmission spectra (see Fig.~\ref{fig5} and extracted bandwidth values in Table~\ref{tab2}), which offers precise frequency selectivity for advanced signal processing applications. 
Moreover, since the SW group velocity is much slower than the speed of light, this enables the construction of delay lines, which are crucial for timing applications during signal processing in modern communication systems. Additionally, the electromagnetic signals, which arrive instantly, can be distinguished from delayed spin-wave signals, enabling the separation of large, potentially damaging signal pulses over time. 
The delay time $T$ can be calculated by dividing the transducer's center-to-center distance between the transducers by the SW group velocity $v_\mathrm{g}$ as follows $T=\frac{G}{v_\mathrm{g}}$. The calculated delay times are in the Table~\ref{tab2}.

Therefore, with the current gap in nano-scale hardware for mobile 5G high-band RF applications and notable advancements in high-quality nanometer-thick LPE YIG film growth \cite{Carsten, Jouseff}, SW technology now presents a promising solution. Below, we outline the advantages and challenges of spin-wave RF technology.

\vspace{0.3cm}
\textbf{\textit{Advantages}}

\begin{itemize}
    \item \textbf{One device for any frequency range.}
    SW-based devices can operate at any frequency range determined by the magnitude of an applied field. Unlike SAW and BAW devices, which must be scaled down to achieve higher frequencies, SW devices maintain the same transducer size and wavelength, regardless of the operating frequency.

    \item \textbf{Industry-ready photolithography-based planar manufacturing.} Although we employed a minimum feature size of $250\,\mathrm{nm}$, our concept can be easily adapted to $500\,\mathrm{nm}$ structures, which are achievable using standard industrial photolithography techniques. The proposed SW technology is planar, facilitating seamless on-chip integration for all 5G frequency bands. Furthermore, unlike SAW/BAW devices, SW isolation is not required as spin waves are always localized within the magnetic material, simplifying the fabrication process. The entire chip size can be less than $1 \times 1\,\mathrm{mm}^2$ even for multi-channel FSLs or filters, as the working areas of the demonstrated devices were below $10 \times 100\,\upmu\mathrm{m}^2$. Consequently, this technology is well-suited for smartphones and other mobile devices. It is also CMOS-compatible when a chiplet approach is employed.

    \item \textbf{3-in-1, energy and space efficient.} SW-based devices enable the integration of multiple functions:-filtering, limiting, and delaying—into a single, passive device. Since no software is required for operation, the SW device can effectively replace three conventional components and, thus, save both space and energy. Additionally, as SW propagation occurs without electron transfer, there is no Joule heating, further enhancing energy efficiency.
 
    \item \textbf{Static and dynamic tunability.}
    The parameters of SW devices can be precisely engineered through various factors, such as the applied magnetic field, SW mode selection, transducer design, magnetic material choice, and the thickness and shape of the magnetic medium. For instance, bandwidth can be adjusted by modifying the transducer design, while the distance between transducers and the thickness of the magnetic film can control the delay time. The threshold power can be tuned through SW mode selection, transducer design, and material choice. Notably, device properties can also be dynamically adjusted on the nanosecond scale \cite{Andrii2}, for example, using variable magnetic or electric (in specially designed hybrid structures \cite{Elfieldcontrol1, Elfieldcontrol2, Elfieldcontrol3, Elfieldcontrol4, Elfieldcontrol5}) fields. This allows ns-fast tuning of the, e.g., FSL or filter passband frequency and bandwidth.
     
    \item \textbf{High signal-to-noise ratio and controllable power threshold.}
    SW-based devices exhibit lower noise levels at high frequencies compared to semiconductor devices. The device is expected to attenuate the parasitic signals with large amplitudes while preserving small signals at different frequencies due to its frequency selectivity, effectively enhancing the signal-to-noise ratio \cite{Adam1, Shukla}. Additionally, SW FSLs have low power thresholds, which are critical for GPS applications. They also improve GPS energy efficiency by eliminating the need for amplifiers for incoming signals. We further propose \cite{patent} that using nano-waveguides instead of planar films leads to the dilution of the magnon mode spectrum, which suppresses multi-magnon scattering processes, providing an additional independent parameter to tune the threshold power. A further reduction in the power threshold might be achieved by utilizing specific points of the spin-wave spectrum with enhanced four-magnon scattering efficiency or, at low frequencies, the three-magnon scattering mechanism \cite{Gurevich-Melkov, LvovBook}, which typically occurs at lower frequencies compared to the four-magnon scattering process.
  
\end{itemize}

\vspace{0.3cm}
\textbf{\textit{Challenges}}

\begin{itemize}
    \item \textbf{Insertion Losses.}
    The smallest insertion losses can be achieved for the transducers with minimal ohmic losses, given by the transducer resistance and its geometry, and with maximum radiation resistance \cite{connely, Budapest, Florian}, determining the efficiency of SW excitation. These two contributions should match the impedance of the system (VNA and coaxial cables, in our case $50\,\Omega$). The radiation resistance is frequency dependent; therefore, the transducers should be optimized for a certain frequency and spin-wave mode. In addition, eddy currents and skin depth may need to be considered, especially at higher operating frequencies. The insertion losses in the studied nanoscale FSL devices currently exceed $20\,\mathrm{dB}$. However, through transducer optimization, these losses can be significantly reduced to as low as approximately $3\,\mathrm{dB}$, which has been proven for lower-frequency band \cite{IL1, IL2, IL3, connely, Budapest, Florian}, but remains a key focus for future research in the middle and upper 5G bands.

    \item \textbf{Need for high magnetic fields.}
    For YIG-based spin-wave RF devices, a magnetic field of $823\,\mathrm{mT}$ is required to operate at 25 GHz (5G high-band) frequencies. While such fields can be achieved using conventional NdFeB magnets, this solution is impractical for most mobile devices unless micromagnets are integrated directly on the chip \cite{patent, Nora}. An alternative approach involves using materials with strong crystallographic anisotropy, such as Perpendicular Magnetic Anisotropy PMA materials \cite{PMA1, PMA2, PMA3} or barium hexaferrites \cite{Hexf1, Hexf2, Hexf3}, which allow operation at frequencies even exceeding 50 GHz with significantly lower magnetic fields in the range of 100~mT.

   \item \textbf{Elimination of the second peak in the transmission spectra.}
    The second peak in the transmission spectra can be effectively eliminated by employing specialized transducer designs. For example, the nonsymmetric input and output transducer, which only partially overlap in their excitation efficiency spectra determining the bandwidth, would minimize the unwanted peaks.
    Additionally, unwanted transmission can be suppressed through the use of magnonic crystals \cite{MC1, MC2, MC3} by inducing anticrossing between the propagating mode and perpendicular standing SW modes \cite{Stopband} or by incorporating other types of SW resonators.
   
    \item \textbf{Temperature Stability.}
    The saturation magnetization $M_S$ is temperature-dependent, and its changes can affect the SW dispersion relation, causing a shift in SW frequencies and influencing the group velocity. YIG has a relatively low Curie temperature (560~K) compared to, e.g., Permalloy (NiFe, 773~K) or Cobalt Iron Boron (CoFeB, 873~K), meaning that $M_S$ of YIG varies with the temperature changes more than in the case of NiFe or CoFeB. Several strategies can be employed to address this challenge, including active temperature stabilization systems and the use of alternative magnetic materials with improved thermal stability.

\end{itemize}

\section{VII. Conclusion} \label{sec:conclusion}
In summary, we have reported on the first study of the nanoscale FSLs based on spin-wave transmission affected by multi-magnon scattering in $97\,\mathrm{nm}$ YIG thin film. The developed FSLs consist of a pair of transducers with the smallest feature size of $250\,\mathrm{nm}$ for SW excitation and detection. We have shown that the same device can work for different 5G frequency bands up to $25\,\mathrm{GHz}$, and it has been tested for DE and BV spin-wave modes. The key device parameters, such as power threshold, power limiting level, insertion losses, and bandwidth, have been extracted and discussed. The lowest power threshold obtained at $4\,\mathrm{GHz}$ and $9\,\mathrm{GHz}$ was $-25.5\pm3.3\,\mathrm{dBm}$ for DE mode, and it was  $0\pm3.5\,\mathrm{dBm}$ at $25\,\mathrm{GHz}$ for BV mode; using the $10\,\upmu\mathrm{m}$ long transducers. The analytical model has been verified to predict power limiting levels, and micromagnetic simulations have been adapted for the quantitative description of the power limiting curves. Possible strategies for further optimization of FSL parameters and the perspective on the spin-wave devices are outlined.

\section*{Data Availability} \label{sec:dataav}
The experimental data supporting this study are publicly available in Phaidra \cite{Phaidra}.
 
\section*{Acknowledgements} \label{sec:acknowledgements}
This research was funded in whole or in part by the Austrian Science Fund (FWF) [10.55776/I4917] and [10.55776/PAT3864023]. For open access purposes, the author has applied a CC BY public copyright license to any author accepted manuscript version arising from this submission.
AC acknowledges the support of the European Research Council (ERC) Proof of Concept Grant 101082020 “5G-Spin” and COST Action CA23136 (CHIROMAG). KL acknowledges the support of FWF ESPRIT Fellowship Grant ESP 526-N TopMag [10.55776/ESP526]. QW was supported by the startup grant of Huazhong University of Science and Technology Grants No. 3034012104. RV acknowledges support from the National Academy of Sciences of Ukraine, project No. 08/01-2024(5). The work of ML was supported by the German Bundesministerium für Wirtschaft und Energie (BMWI) under Grant No. 49MF180119. We acknowledge CzechNanoLab Research Infrastructure supported by MEYS CR (LM2023051). We are grateful to Prof. P. Pirro for fruitful discussions on the multi-magnon scattering phenomena. We express our gratitude to Dr. Vojtěch Švarc for his valuable advice on sample fabrication.
We thank Barbora and Sabri Koraltan for their hospitality and facilitation of valuable discussions.

\end{document}